\documentstyle[preprint,epsf,aps,floats]{revtex}
\begin{document}
\tightenlines

\title{Measurement of Dijet Angular Distributions 
and Search for Quark Compositeness}

\author{                                                                      
B.~Abbott,$^{28}$                                                             
M.~Abolins,$^{25}$                                                            
B.S.~Acharya,$^{43}$                                                          
I.~Adam,$^{12}$                                                                
D.L.~Adams,$^{37}$                                                            
M.~Adams,$^{17}$                                                              
S.~Ahn,$^{14}$                                                                
H.~Aihara,$^{22}$                                                             
G.A.~Alves,$^{10}$                                                            
E.~Amidi,$^{29}$
% ...
\\                                                                            
\vskip 0.50cm                                                                 
\centerline{(D\O\ Collaboration)}                                             
\vskip 0.50cm                                                                 
}                                                                             
\address{                                                                     
\centerline{$^{1}$Universidad de los Andes, Bogot\'{a}, Colombia}             
\centerline{$^{2}$University of Arizona, Tucson, Arizona 85721}               
\centerline{$^{3}$Boston University, Boston, Massachusetts 02215}             
\centerline{$^{4}$Brookhaven National Laboratory, Upton, New York 11973}      
\centerline{$^{5}$Brown University, Providence, Rhode Island 02912}           
\centerline{$^{6}$Universidad de Buenos Aires, Buenos Aires, Argentina}       
\centerline{$^{7}$University of California, Davis, California 95616}          
\centerline{$^{8}$University of California, Irvine, California 92697}         
\centerline{$^{9}$University of California, Riverside, California 92521}      
\centerline{$^{10}$LAFEX, Centro Brasileiro de Pesquisas F{\'\i}sicas,          
                  Rio de Janeiro, Brazil}                                     
\centerline{$^{11}$CINVESTAV, Mexico City, Mexico}                            
\centerline{$^{12}$Columbia University, New York, New York 10027}             
\centerline{$^{13}$Delhi University, Delhi, India 110007}                     
\centerline{$^{14}$Fermi National Accelerator Laboratory, Batavia,            
                   Illinois 60510}                                            
\centerline{$^{15}$Florida State University, Tallahassee, Florida 32306}      
\centerline{$^{16}$University of Hawaii, Honolulu, Hawaii 96822}              
\centerline{$^{17}$University of Illinois at Chicago, Chicago,                
                   Illinois 60607}                                            
\centerline{$^{18}$Indiana University, Bloomington, Indiana 47405}            
\centerline{$^{19}$Iowa State University, Ames, Iowa 50011}                   
\centerline{$^{20}$Korea University, Seoul, Korea}                            
\centerline{$^{21}$Kyungsung University, Pusan, Korea}                        
\centerline{$^{22}$Lawrence Berkeley National Laboratory and University of    
                   California, Berkeley, California 94720}                    
\centerline{$^{23}$University of Maryland, College Park, Maryland 20742}      
\centerline{$^{24}$University of Michigan, Ann Arbor, Michigan 48109}         
\centerline{$^{25}$Michigan State University, East Lansing, Michigan 48824}   
\centerline{$^{26}$Moscow State University, Moscow, Russia}                   
\centerline{$^{27}$University of Nebraska, Lincoln, Nebraska 68588}           
\centerline{$^{28}$New York University, New York, New York 10003}             
\centerline{$^{29}$Northeastern University, Boston, Massachusetts 02115}      
\centerline{$^{30}$Northern Illinois University, DeKalb, Illinois 60115}      
\centerline{$^{31}$Northwestern University, Evanston, Illinois 60208}         
\centerline{$^{32}$University of Notre Dame, Notre Dame, Indiana 46556}       
\centerline{$^{33}$University of Oklahoma, Norman, Oklahoma 73019}            
\centerline{$^{34}$University of Panjab, Chandigarh 16-00-14, India}          
\centerline{$^{35}$Institute for High Energy Physics, 142-284 Protvino,       
                   Russia}                                                    
\centerline{$^{36}$Purdue University, West Lafayette, Indiana 47907}          
\centerline{$^{37}$Rice University, Houston, Texas 77005}                     
\centerline{$^{38}$Universidade do Estado do Rio de Janeiro, Brazil}          
\centerline{$^{39}$University of Rochester, Rochester, New York 14627}        
\centerline{$^{40}$CEA, DAPNIA/Service de Physique des Particules,            
                   CE-SACLAY, Gif-sur-Yvette, France}                         
\centerline{$^{41}$Seoul National University, Seoul, Korea}                   
\centerline{$^{42}$State University of New York, Stony Brook,                 
                   New York 11794}                                            
\centerline{$^{43}$Tata Institute of Fundamental Research,                    
                   Colaba, Mumbai 400005, India}                              
\centerline{$^{44}$University of Texas, Arlington, Texas 76019}               
\centerline{$^{45}$Texas A\&M University, College Station, Texas 77843}       
}                                                                             

\date{\today}

\maketitle
                                     
\begin{abstract}
We have measured the dijet angular distribution in
$\sqrt{s}$=1.8 TeV $p\bar{p}$ collisions using the D\O\ detector.  
Order $\alpha^{3}_{s}$ QCD predictions
are in good agreement with the data.  
At 95\% confidence the data exclude models of quark compositeness 
in which the
contact interaction scale is below 2 TeV.
\end{abstract}

\vspace {.25in}

  Dijet final states in $p \bar{p}$ collisions
can be produced through quark-quark, quark-gluon and gluon-gluon
interactions.  
The angular distributions produced by these processes as
predicted by theory are similar.
Therefore the dijet angular distribution is insensitive to the
relative weighting of the individual hard scattering process and thus
is insulated from uncertainties in the parton distribution 
functions (pdf's).
Quantum chromodynamics 
(QCD) parton-parton scattering processes, which are mainly 
{\it t}-channel exchanges,
produce dijet angular distributions peaked at small
center of mass scattering angles, 
while many processes containing new physics 
are more isotropic.  Thus, the dijet angular distribution 
provides an
excellent test of QCD and a means of searching for new physics such
as quark compositeness.
We have measured the dijet angular distribution 
with greater precision
over a larger dijet invariant mass range and a larger
angular range than previous measurements.
Next-to-leading order (NLO) QCD predictions are 
available \cite{EKS,theory}
and comparisons can be made between the data and both
leading order (LO) and NLO predictions.  

The value of the mass scale, $\Lambda$, characterizes the strength
of the quark substructure binding interactions and the physical size
of the composite states. 
In the regime where \mbox{$\Lambda$ $\gg$ $\sqrt{\hat{s}}$}
is valid, the quarks appear almost pointlike and any quark
substructure coupling can be approximated by a four-Fermi interaction.  
With this approximation, the
effective Lagrangian for a flavor diagonal definite chirality 
current is 
\cite{compos,compos2}: 
\mbox{$\cal L$ =$A$(2$\pi/ \Lambda^{2})(\bar{q}_{H}\gamma^{\mu}q_{H})
(\bar{q}_{H}\gamma_{\mu}q_{H})$}
where $A=\pm 1$, and $H$= $L$, $R$ for left or right handed quarks.  
While this is not the only possible contact
interaction, it is the only one for which calculations are currently 
available.
Since the sign of $A$ is a priori undetermined, limits
for constructive interference ($A$=$-1$) and destructive 
interference ($A$=$+1$) are
presented. 
Previously published results from CDF \cite{CDF} on dijet angular 
distributions have 
been compared to the same model in which all quarks are composite, 
yielding 
95\% confidence limits $\Lambda^{+}>1.8$ TeV and $\Lambda^{-}>1.6$ 
TeV on the interaction scales.   

The D\O\ detector, described in detail elsewhere \cite{det}, measures
jets using uranium-liquid argon sampling calorimeters that
provide uniform
and hermetic coverage over a large range of pseudorapidity 
($|\eta|$ $\le$ 4).  
Typical
transverse segmentation is 0.1 $\times$ 0.1 in $\eta$ $\times$
$\phi$, where $\phi$ is the azimuthal angle.

The data are from the 93 $\rm pb^{-1}$ sample recorded during 
the 1994-1995 Tevatron
run.
Events are selected using a multi-level trigger.
The first level requires an inelastic collision by demanding
the coincidence of two hodoscopes on either side of the 
interaction region.
In the second level, jet candidates are selected using an array
of 40 calorimeter trigger towers 0.8 $\times$ 1.6 in $\eta \times \phi$, 
covering $|\eta|$ $<$ 4.
Four different trigger criteria are defined, each requiring
a single trigger tower above a different transverse energy 
($E_T$) threshold.
The final level, an online software trigger,
selects events with a jet above a preset threshold.  
The $E_{T}$ thresholds at which the triggers are $>$ 98\% 
efficient for the $\eta$ 
coverage used in this analysis are 55, 90, 120 and 175 GeV.

Jets are reconstructed using a fixed cone algorithm with 
radius ${\cal R}$=
$\sqrt{\Delta \eta^{2} + \Delta \phi^{2}}$ = 0.7.  Calorimeter towers
with $E_{T}$
greater than 1.0 GeV are used as seed towers for jet finding
\cite{jetfinding}.  
Jet $E_{T}$ is defined as the sum of the $E_{T}$ in the towers with 
${\cal R}$ $<$ 0.7 
from the seed tower and
a new $E_{T}$ 
weighted center of the jet is calculated.  This process is repeated until
the jet center is stable.  When one jet overlaps another, they are
merged into a single jet if they share more than 50\% of the $E_{T}$ of the
lower $E_{T}$ jet.  Otherwise, they are split into two distinct jets.

Jet energy calibration is performed in a multi-step process \cite{resp}.
First, the
electromagnetic energy scale in the central calorimeter is determined by 
scaling the energies of electrons from $Z$ boson decays so that
the corrected $Z$ mass agrees with the value measured at
LEP \cite{lep}.
Next, the jet response of the central calorimeter is measured using
momentum conservation in a sample of photon + jet events.
After determining the jet response for the central calorimeter as a 
function of
jet energy, the
relative $\eta$ dependent jet response is measured using
both photon + jet and dijet events.  
One jet (photon) is required to be central and the jet
response is measured as a function of the $\eta$ of the other jet.
Jets are also corrected for out-of-cone showering losses, underlying event, 
multiple $p\bar{p}$ interactions, and effects of uranium noise.

Quality cuts are required for the two leading $E_{T}$ jets in each event.
These cuts eliminate spurious jets that arise from noisy calorimeter cells, 
cosmic rays, and accelerator losses. The efficiencies for these cuts 
are $E_{T}$ 
and $\eta$ dependent
and vary between 90\% and 97\%.

The dijet invariant mass and kinematic variables of an event are defined 
using the two 
highest $E_{T}$ jets.  The center of mass scattering angle, $\theta^{*}$, 
and the
longitudinal boost, $\eta_{\rm boost}$, are expressed in terms 
of the pseudorapidities of these
two jets: $\eta_{\rm boost}$=$\frac{1}{2}$($\eta_{1}$+$\eta_{2}$) 
and $\cos{\theta^{*}}$=
$\tanh{\eta^{*}}$, where $\eta^{*}$=$\frac{1}{2}$($\eta_{1}-\eta_{2}$).  
The dominance
of spin-1 gluon exchange gives the angular distribution characteristic of
Rutherford scattering: d{\it N}/d$\cos{\theta^{*}}$ $\propto$ 
1/$\sin^{4}({\theta^{*}}/2$).
To facilitate a comparison with theory, the angular distributions
are measured as a function of 
\mbox{$\chi$=$e^{2|\eta^{*}|}$=
(1+$|\cos{\theta^{*}}|)$/(1$-|\cos{\theta^{*}}|)$}.
This definition translates large values of $\theta^{*}$ to small values of
$\chi$ (e.g. $\theta^{*}=90^{\circ}$ $\leftrightarrow$ $\chi$=1).
For Rutherford scattering, d{\it N}/d$\chi$ is independent of $\chi$.

The data are selected using a dijet invariant mass ($M$) threshold 
and restricting
the kinematic cuts in order to limit the jets to regions of 
uniform acceptance
\cite{kathy}.
The dijet invariant mass is calculated assuming massless jets and using 
the expression:
\mbox{$M^{2}=2E_{T1}E_{T2}
(\cosh{(\eta_{1}-\eta_{2})}-\cos{(\phi_{1}-\phi_{2})})$}.
Table \ref{TABLE:range} shows the average dijet invariant mass, 
maximum $\chi$ measured, and
the number of events for each of four mass ranges.
Both $|\eta_{1}|$ and $|\eta_{2}|$ are required
to be less than 3. 
To maintain uniform acceptance, we also require 
$|\eta_{\rm boost}|$ $<$ 1.5.

\begin{table}
\begin{center}
\caption{The average mass, maximum $\chi$ measured, and
the number of events after applying all kinematic cuts.}
\begin{tabular}{ccccc}
\multicolumn{1}{c}{Trigger $E_{T}$} &
\multicolumn{1}{c}{Mass} &
\multicolumn{1}{c}{Average} &
\multicolumn{1}{c}{$\chi_{\rm max}$} &
\multicolumn{1}{c}{Number} \\
\multicolumn{1}{c}{Threshold} &
\multicolumn{1}{c}{Range} &
\multicolumn{1}{c}{Mass} &
\multicolumn{1}{c}{ } &
\multicolumn{1}{c}{of Events} \\
\multicolumn{1}{c}{(GeV)} &
\multicolumn{1}{c}{(GeV/$\rm c^{2}$)} &
\multicolumn{1}{c}{(GeV/$\rm c^{2}$)} &
\multicolumn{1}{c}{ } &
\multicolumn{1}{c}{ } \\
\hline
55  & 260-425 & 302 & 20 & 4621 \\
90  & 425-475 & 447 & 20 & 1573 \\
120  & 475-635 & 524 & 13 & 8789 \\
175  & $>$635    & 700 & 11 & 1074 \\
\end{tabular}
\label{TABLE:range}
\end{center}
\end{table}

\begin{figure}
\begin{center}
\begin{tabular}{c}
   \epsfxsize = 9.0 cm \epsffile{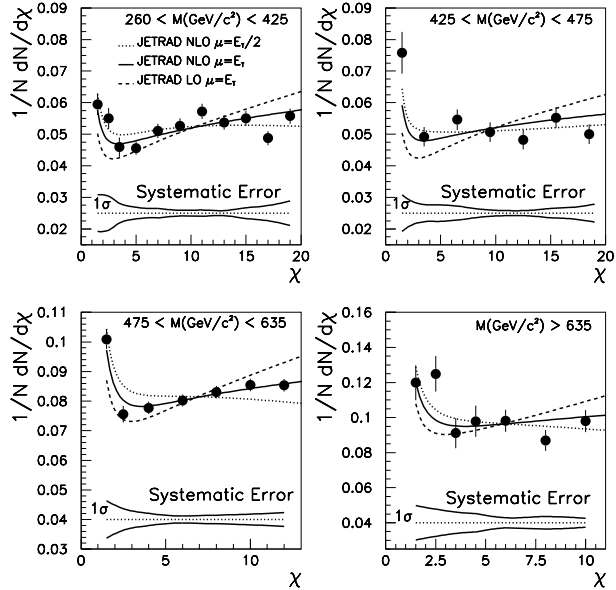}
\end{tabular}
\caption{Dijet angular distributions for D\O\ data (points)
compared to {\sc jetrad} for LO (dashed line) and NLO predictions
with renormalization/factorization scale $\mu$=$E_{T}$(solid line).  
The data are also
compared to {\sc jetrad} NLO predictions 
with $\mu$=$E_{T}$/2 (dotted line).
The errors on the data points are statistical only. The band at the bottom
represents the correlated $\pm$ 1 $\sigma$ systematic uncertainty.}
\label{FIG:data_vs_lo_nlo}
\end{center}
\end{figure}

Table \ref{TABLE:numbers} shows the dijet angular distribution, 
(100/$N$)(d$N$/d$\chi$), in
bins of $\Delta \chi$ = 1 with
its statistical error in the four mass bins.
The {\sc jetrad} program \cite{theory} is used to determine the LO
and NLO QCD predictions.
The jets at NLO
are found using the standard \cite{snowmass} jet definition which 
combines two partons
into a single jet if they are both within ${\cal R}$=0.7 of their 
$E_{T}$ weighted center.
We require that two    
partons also be closer than ${\cal R}_{\rm sep}$ $\times$ 0.7 with 
${\cal R}_{\rm sep}$=1.3 
\cite{rsep,Klasen}.  
Figure \ref{FIG:data_vs_lo_nlo} shows the dijet angular 
distributions normalized
to unit area compared to three different theoretical predictions.
The dashed and solid curves show the LO and NLO predictions for
a single choice of renormalization/factorization
scale, $\mu$=$E_{T}$ of the leading jet.   
The dotted curve shows the effects of changing 
the scale to $\mu$=$E_{T}/2$ for the NLO predictions.
The LO predictions are insensitive to the renormalization scale,
so only one scale is shown.
With the large value of $\chi_{\rm max}$, the effects of higher order
QCD become apparent.
The theoretical predictions are clearly sensitive to the order of the 
calculation and to the renormalization scale.  
The NLO predictions are seen to be in better
agreement with the data than the LO calculations, especially for
large $\chi$.

\begin{table}
\begin{center}
\caption{Dijet angular distribution (100/$N$)(d$N$/d$\chi$) with 
statistical uncertainties for
the four mass bins (GeV/$\rm c^{2}$).}
\begin{tabular}{ccccc}
\multicolumn{1}{c}{$\chi$} &
\multicolumn{1}{c}{260$<M<$425} &
\multicolumn{1}{c}{425$<M<$475} &
\multicolumn{1}{c}{475$<M<$635} &
\multicolumn{1}{c}{$M>$635} \\
\hline
1.5  & 5.94 $\pm$ 0.36 & 7.58 $\pm$ 0.69 & 10.1 $\pm$ 0.34 & 12.0  
$\pm$  1.04  \\
2.5  & 5.50 $\pm$ 0.35 & 4.26 $\pm$ 0.52 & 7.56 $\pm$ 0.30 & 12.5  
$\pm$  1.06  \\
3.5  & 4.59 $\pm$ 0.32 & 4.96 $\pm$ 0.57 & 7.83 $\pm$ 0.30 & 9.11  
$\pm$  0.91  \\
4.5  & 4.57 $\pm$ 0.32 & 5.54 $\pm$ 0.59 & 7.71 $\pm$ 0.30 & 9.79  
$\pm$  0.95  \\
5.5  & 4.56 $\pm$ 0.32 & 5.29 $\pm$ 0.58 & 7.87 $\pm$ 0.31 & 10.1  
$\pm$  0.97  \\
6.5  & 5.10 $\pm$ 0.33 & 6.26 $\pm$ 0.64 & 8.17 $\pm$ 0.31 & 9.58  
$\pm$  0.95  \\
7.5  & 5.10 $\pm$ 0.33 & 4.83 $\pm$ 0.56 & 8.70 $\pm$ 0.32 & 9.30  
$\pm$  0.94  \\
8.5  & 5.61 $\pm$ 0.35 & 4.40 $\pm$ 0.53 & 7.91 $\pm$ 0.31 & 8.08  
$\pm$  0.88  \\
9.5  & 4.93 $\pm$ 0.33 & 5.60 $\pm$ 0.60 & 8.46 $\pm$ 0.32 & 8.96  
$\pm$  0.92  \\
10.5 & 6.04 $\pm$ 0.36 & 5.21 $\pm$ 0.58 & 8.62 $\pm$ 0.32 & 10.6  
$\pm$  1.01  \\
11.5 & 5.40 $\pm$ 0.34 & 4.30 $\pm$ 0.53 & 8.38 $\pm$ 0.32 &  \\
12.5 & 5.33 $\pm$ 0.34 & 4.75 $\pm$ 0.55 & 8.69 $\pm$ 0.32 &  \\
13.5 & 5.41 $\pm$ 0.34 & 5.43 $\pm$ 0.58 &  &  \\
14.5 & 5.40 $\pm$ 0.34 & 5.69 $\pm$ 0.60 &  &  \\
15.5 & 5.60 $\pm$ 0.35 & 6.18 $\pm$ 0.63 &  &  \\
16.5 & 4.81 $\pm$ 0.32 & 4.70 $\pm$ 0.55 &  &  \\
17.5 & 4.95 $\pm$ 0.33 & 4.83 $\pm$ 0.55 &  &  \\
18.5 & 5.78 $\pm$ 0.35 & 5.01 $\pm$ 0.56 &  &  \\
19.5 & 5.37 $\pm$ 0.34 & 5.17 $\pm$ 0.57 &  &  \\
\end{tabular}
\label{TABLE:numbers}
\end{center}
\end{table}

The dominant source of error on the angular 
distribution is the uncertainty in the $\eta$ dependence of the
calorimeter energy scale, which is found to be
less than 2\%.
Other systematic uncertainties, including
$\eta$ biases in jet reconstruction, multiple $p\bar{p}$ interactions,
$\eta$ dependent jet quality cut efficiencies, and effects of
calorimeter $\eta$ and $E_{T}$ smearing are small.
Because the data distributions are normalized to unit area, 
uncertainties in the absolute jet energy scale are
minimal.  All systematic uncertainties added in quadrature
are shown as a band at the bottom of Fig. 1. 
The effects of a different pdf were examined by replacing the default
CTEQ3M \cite{3m} with CTEQ2MS \cite{2ms}.
The calculated angular distribution is insensitive to this change.  

\begin{figure}
\begin{center}
\begin{tabular}{c}
   \epsfxsize = 9.0 cm \epsffile{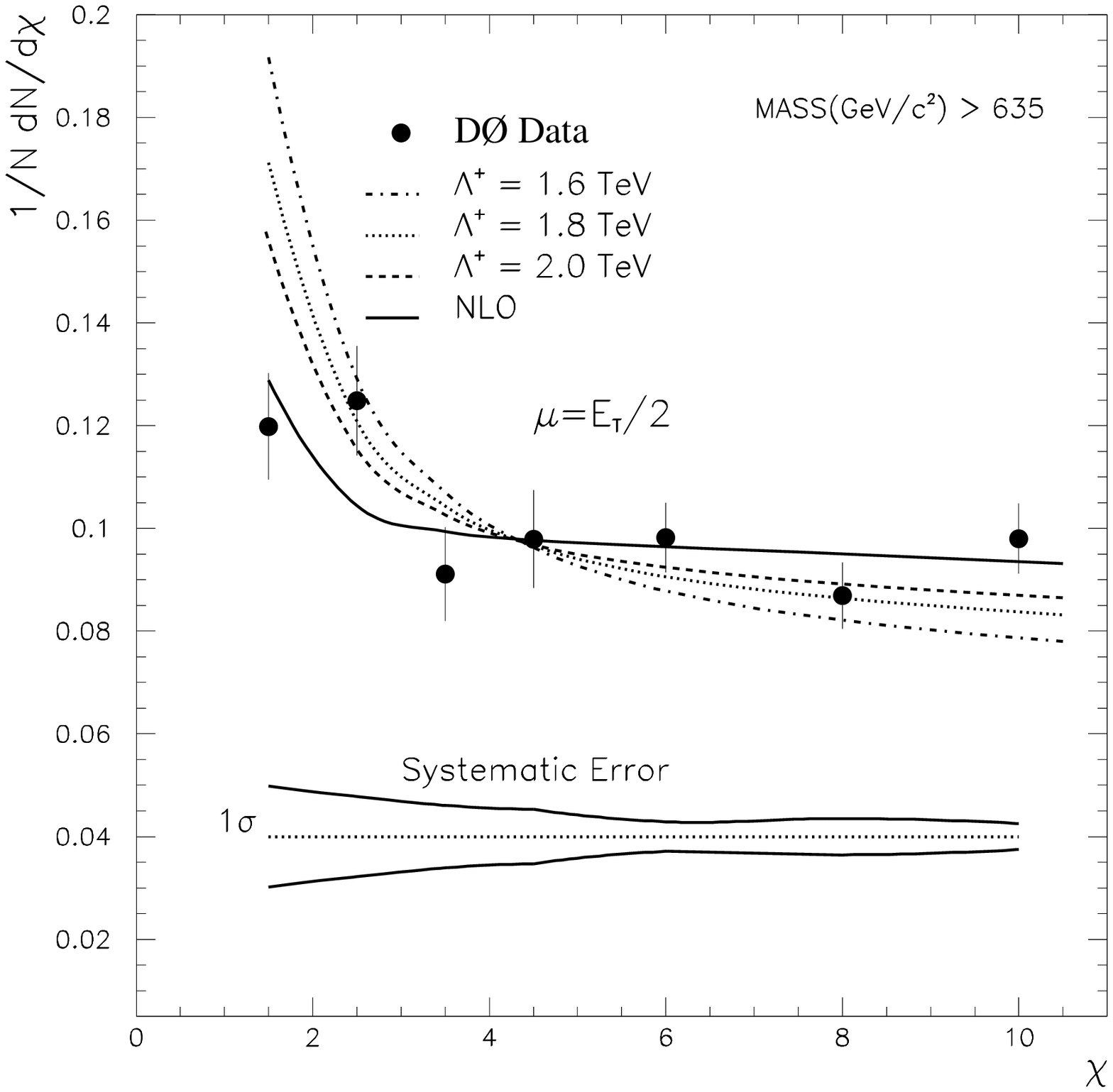}
\end{tabular}
\caption{Data compared to theory for different compositeness scales.  See
text for how compositeness is calculated for NLO predictions.  
The errors on the points
are statistical and the band represents the correlated $\pm$ 1 $\sigma$
systematic uncertainty.}
\label{FIG:comp}
\end{center}
\end{figure}

Because the currently available NLO calculations 
do not implement the effects of both QCD and quark substructure, 
possible effects of quark compositeness are determined using
a LO simulation \cite{compos2}.  
The ratio of the LO predictions with compositeness 
to the LO predictions with no compositeness is used to scale 
the NLO calculations.
Figure \ref{FIG:comp}
shows the dijet angular distribution for events with $M$ $>$ 
635 GeV/$\rm c^{2}$ compared to theory for different values of
the compositeness scale, $\Lambda^{+}$.  The largest dijet 
invariant mass bin is shown because
the effects of quark compositeness become more pronounced with increasing
dijet invariant mass.  

\begin{figure}
\begin{center}
\begin{tabular}{c}
   \epsfxsize = 9.0 cm \epsffile{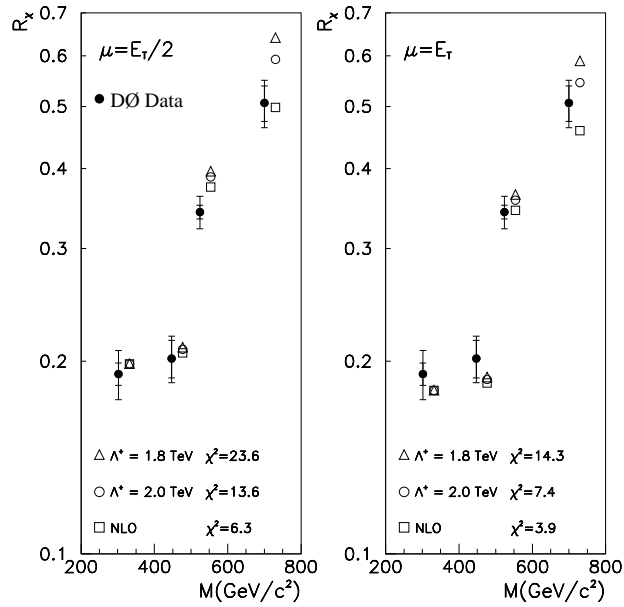}
\end{tabular}
\caption{$R_{\rm \chi}$ as a function of dijet invariant mass for 
two different
renormalization scales.  See text for how compositeness is calculated
for NLO predictions. 
The inner error bars are the statistical uncertainties and the 
outer error bars
include the statistical and systematic uncertainties added in quadrature.  
The $\chi^{2}$ values for the four degrees
of freedom are shown for the different values of the compositeness scale.
The D\O\ data are plotted
at the average mass for each mass range.  The NLO points are offset 
in mass to allow the
data points to be seen.}
\label{FIG:R}
\end{center}
\end{figure}

To obtain a compositeness limit, we constructed the variable
$R_{\rm \chi}$, the ratio of the
number of events with $\chi < 4$ to the number of events with 
$4 < \chi < \chi_{\rm max}$.
The value $\chi$ = 4 is chosen to optimize the sensitivity to
quark compositeness. 
Because the angular distribution of jets arising from 
contact interactions is expected to be
more isotropic than that for QCD interactions, contact interactions
will produce more events at small
$\chi$ than QCD and therefore will have a larger value of $R_{\rm \chi}$.
Table \ref{TABLE:R} shows the experimental ratio $R_{\rm \chi}$
for the different mass ranges with their statistical
and their systematic uncertainties, which are fully correlated in mass.  
Figure \ref{FIG:R} exhibits $R_{\rm \chi}$ as a function of $M$ 
for two different renormalization scales along with 
the theoretical predictions for different compositeness scales.
Note that the two largest dijet
invariant mass bins have a lower $\chi_{\rm max}$ value 
(Table \ref{TABLE:range}), and thus a higher
value of $R_{\rm \chi}$ is expected independent of compositeness assumptions.
Also shown in Fig. \ref{FIG:R} are the $\chi^{2}$ values for the four degrees
of freedom for different values of the compositeness scale.
The $\chi^{2}$ is defined
as $\chi^{2}$=$\sum_{i,j} \delta_{i}V^{-1}_{ij}\delta_{j}$, where
$\delta_{i}$ is the difference between data and theory in each mass bin $i$.
The covariance matrix, $V$, is defined as 
$V_{ii}$=$\sigma^{2}_{i}$(stat.)+$\sigma^{2}_{i}$(syst.),
$V_{ij}=\sigma_{i}$(syst.)$\sigma_{j}$(syst.), for $i \neq j$.
For both renormalization scales, the data prefer a model 
without quark compositeness.
The data are better fit with $\mu$=$E_{T}$.   

We employed a Bayesian technique \cite{Bayes} to obtain a
compositeness scale limit from our data, using
a Gaussian likelihood function, $P(R_{\rm \chi}|\xi)$, 
for $R_{\rm \chi}$ as a function of dijet 
invariant mass.
The compositeness limit depends on the choice of the
prior probability distribution, $P(\xi)$.  Motivated by the
form of the Lagrangian, $P(\xi)$ is assumed
to be either flat in $\xi$=1/$\Lambda^2$ or $\xi$=1/$\Lambda^4$.
Since the dijet angular distribution at NLO is sensitive to 
the renormalization scale, each renormalization scale is treated as a 
different theory.  
The likelihood function has the form
$P(R_{\rm \chi}|\xi) = e^{-\chi^{2}/2}$.
To determine the 95\% confidence limit in $\Lambda$, 
a limit in $\xi$ is first
calculated by requiring that 
$Q(\xi)$ =$\int_{0}^{\xi}P(R_{\rm \chi}|\xi') 
P(\xi')d\xi'$ = 0.95 of $Q(\infty)$.   
The limit in $\xi$ is then transformed back into a limit in $\Lambda$. 
Table \ref{TABLE:limits} shows the 95\%
confidence limits for the compositeness scale 
obtained for different choices of models
using a prior probability distribution
which is flat in 1/$\Lambda^2$. 

If we vary the models to include 
constructive interference ($\Lambda^{-}$), or require only up and down 
quarks to be composite ($\Lambda_{ud}$), the 95\% confidence limits for 
the compositeness scale change by 0.1 TeV.
If the prior probability distribution is assumed
to be flat in  1/$\Lambda^4$, the 95\% confidence limits are reduced by 
approximately
6\%. 
These limits are valid for either pure left- or right-handed contact 
interactions. 
Unlike the earlier measurement\cite{CDF} using $\chi_{\rm max}$=5, 
the large $\chi$
explored here gives greater sensitivity to compositeness terms 
with constructive
interference than for destructive interference.

\begin{table}
\begin{center}
\caption{Values of $R_{\rm \chi}$ with statistical and
fully correlated systematic uncertainties.}
\begin{tabular}{cccc}
\multicolumn{1}{c}{Mass Range (GeV/$\rm c^{2}$}) &
\multicolumn{1}{c}{$R_{\rm \chi}$} &
\multicolumn{1}{c}{Stat. Error} &
\multicolumn{1}{c}{Syst. Error} \\
\hline
260-425  & 0.191 & 0.0077 & 0.015 \\
425-475  & 0.202 & 0.0136 & 0.010 \\
475-635  & 0.342 & 0.0085 & 0.018 \\
$>$ 635  & 0.506 & 0.0324 & 0.028 \\
\end{tabular}
\label{TABLE:R}
\end{center}
\end{table}

\begin{table}
\begin{center}
\caption{The 95\% confidence limits for the
left-handed contact compositeness scale for different
models.  The prior probability distribution is assumed
to be flat in 1/$\Lambda^2$.}
\begin{tabular}{ccc}
\multicolumn{1}{c}{Compositeness scale} &
\multicolumn{1}{c}{$\mu$=$E_{T}$} &
\multicolumn{1}{c}{$\mu$=$E_{T}$/2} \\
\hline
$\Lambda^{+}$  & 2.1 TeV & 2.3 TeV \\
$\Lambda^{-}$  & 2.2 TeV & 2.4 TeV \\
$\Lambda_{ud}^{-}$ & 2.0 TeV & 2.2 TeV \\
\end{tabular}
\label{TABLE:limits}
\end{center}
\end{table}

In conclusion, we have measured the dijet angular distribution 
with greater precision over
a larger dijet invariant mass range and a larger
angular range than previous measurements.
The data distributions are in good agreement with NLO QCD predictions.
The compositeness limit depends on the choice of the 
renormalization/factorization
scale, the model of compositeness, and the choice of the 
prior probability function. 
We have presented 
compositeness limits for models with left-handed contact
interference.  
With 95\% confidence, 
the interaction scales $\Lambda^{+}$, $\Lambda^{-}$, 
and $\Lambda_{ud}^{-}$ 
all exceed 2.0 TeV.

We thank R. Harris for the use of his program based on
Ref. \cite{compos2}.
We also thank the staffs at Fermilab and collaborating 
institutions for their
contributions to this work, and acknowledge support from the
Department of Energy and National Science Foundation (U.S.A.),
Commissariat  \` a L'Energie Atomique (France),
State Committee for Science and Technology and Ministry for Atomic
   Energy (Russia),
CNPq (Brazil),
Departments of Atomic Energy and Science and Education (India),
Colciencias (Colombia),
CONACyT (Mexico),
Ministry of Education and KOSEF (Korea),
and CONICET and UBACyT (Argentina).


\begin{thebibliography}{99}
\bibitem[*]{beijing}
Visitor from IHEP, Beijing, China.

\bibitem[\dag]{ecuador}
Visitor from Univ. San Francisco de Quito, Ecuador.

\vskip 0.25cm


\bibitem{EKS}  S.D. Ellis, Z. Kunszt, and D.E. Soper, 
Phys. Rev. Lett. {\bf 64}, 2121 (1990).

\bibitem{theory}
   W.T. Giele, E.W.N. Glover and D.A. Kosower,
   Nucl. Phys. B {\bf 403}, 633 (1993).

\bibitem{compos}  E. Eichten, K. Lane, and M. Peskin, 
Phys. Rev. Lett. {\bf 50}, 811 (1983).

\bibitem{compos2} K. Lane, BUHEP-96-8, hep-ph/9605257 (1996).

\bibitem{CDF}  CDF Collaboration, F. Abe  {\em et al.}, 
              Phys. Rev. Lett. {\bf 77}, 5336 (1996). 
              Erratum - ibid. {\bf 78}, 4307 (1997).

\bibitem{det} D\O\ Collaboration, S. Abachi {\em et al.}, 
       Nucl. Instrum. and Methods A {\bf 338}, 185 (1994). 
\bibitem{jetfinding} D\O\ Collaboration, S. Abachi {\em et al.},   
Phys. Lett. B {\bf 357}, 500 (1995).
\bibitem{resp} D\O\ Collaboration, R. Kehoe, to be published in 
Proc. ${6^{th}}$ International Conf.
  on Calorimetry in High Energy Physics, Frascati (1996),
Fermilab-Conf-96/284-E.
\bibitem{lep} Particle Data Group, ``Review of Particle Properties",
Phys. Rev D{\bf 50}, 1336 (1994).
\bibitem{kathy}  Mary K. Fatyga, Ph.D. dissertation, 
University of Rochester, 1996 (unpublished).
\bibitem{snowmass} J. Huth {\em et al.}, in proceedings of 
{\em Research Directions
  for the Decade, Snowmass 1990}, edited by  E.L. Berger (World
  Scientific, Singapore, 1992).
\bibitem{rsep} S.D. Ellis, Z. Kunzst, and D.E. Soper, 
Phys. Rev. Lett. {\bf69} ,3615 (1992).
\bibitem{Klasen}  M. Klasen and G. Kramer, hep-ph/9701247 (January 1997).
\bibitem{3m} H.L. Lai {\em et al.}, (CTEQ Collaboration) 
Phys. Rev. D {\bf51}, 4763 (1995).
\bibitem{2ms} J. Botts, {\em et al.}, (CTEQ Collaboration)
    Phys. Lett. B {\bf 304}, 159 (1993).
\bibitem{Bayes} H. Jeffreys, Theory of Probability, 
Clarendon Press, Oxford (1939, revised 1988).
\end{thebibliography}
\end{document}